\documentclass[aps,prl,twocolumn,reprint,superscriptaddress,amsmath,amssymb]{revtex4-2}
\usepackage{graphicx}
\usepackage{xcolor}
\newcommand{\avg}[1]{\left\langle #1 \right\rangle}

\begin{document}

\title{Teleported free energy simulations}
\author{Davide Carbone}
\affiliation{Centre d'\'Etudes et de Recherche en Informatique et Communications (CEDRIC),\\ Conservatoire National des Arts et M\'etiers (Cnam), 292 rue Saint-Martin, 75003 Paris, France}
\email{davide.carbone@lecnam.net}

\begin{abstract}
Escorted free energy simulations reduce dissipation with a flow field that follows the equilibrium distribution. A continuous flow, however, cannot transfer probability between metastable basins. We show that any escort then faces a variance floor set by the change in basin weights, and lift it by adding teleportation: jumps at a rate proportional to the driving rate. We derive the corresponding nonequilibrium identity, the zero-dissipation condition, and the minimal number of jumps it requires, and illustrate the gain on an exactly solvable model and a tilted double well.
\end{abstract}

\maketitle

Computing the free energy difference $\Delta F=F_B-F_A$ between two equilibrium states of a system, held at the same temperature $\beta^{-1}$ but at different values $\lambda=A,B$ of an external parameter, is a routine and yet demanding task of molecular simulation. Alongside equilibrium methods such as thermodynamic integration and free energy perturbation, nonequilibrium approaches have become a standard tool: one repeatedly drives $\lambda$ from $A$ to $B$ in a finite time $\tau$, starting from equilibrium samples, and estimates $\Delta F$ through the nonequilibrium work relation~\cite{Jarzynski1997,Crooks1999}
\begin{equation}
e^{-\beta\Delta F}=\avg{e^{-\beta W}} ,
\label{eq:JE}
\end{equation}
where $W$ is the work performed during a single realization and the angular brackets denote an average over realizations.

Equation~(\ref{eq:JE}) holds at any switching rate, but its usefulness depends on how fast the exponential average converges. The relevant scale is the dissipated work $\avg{W}-\Delta F\ge0$. When the system falls out of step with the instantaneous equilibrium distribution, the average becomes dominated by rare realizations with atypically low work, and the number of trajectories needed grows rapidly, typically exponentially, with the dissipated work~\cite{Jarzynski2006,VJ2009}. Vaikuntanathan and Jarzynski showed that this lag can be attacked at its source~\cite{VJ2008}. Consider a system with Hamiltonian $H(\mathbf z,\lambda)$ and equilibrium distribution $p_{\rm eq}=e^{-\beta H}/Z$, where $Z(\lambda)=\int d\mathbf z\,e^{-\beta H}=e^{-\beta F}$; we write $Z_A=Z(A)$. One adds to the equations of motion an artificial flow field $\mathbf u$ that couples the system coordinates $\mathbf z$ directly to changes in $\lambda$,
\begin{equation}
\dot{\mathbf z}=\tilde{\mathbf v}(\mathbf z,\lambda)+\dot\lambda\,\mathbf u(\mathbf z,\lambda),
\label{eq:escort}
\end{equation}
and corrects the work with the corresponding Jacobian term. The result is an exact identity of the same form as Eq.~(\ref{eq:JE}). If $\mathbf u$ moves the system exactly as the equilibrium distribution moves, the lag disappears and every trajectory returns $W=\Delta F$. The idea has since been extended to discrete-time maps~\cite{VJ2011}, to learned flows~\cite{Wirnsberger2020,NETS2024}, and to the virtual escorting of unmodified trajectories~\cite{LeeJarzynski2026}.

An escort, however, is a smooth flow, and a smooth flow can only rearrange probability continuously. This is harmless as long as $\lambda$ merely deforms the equilibrium distribution, shifting or reshaping its peaks. It becomes a genuine obstruction when $\lambda$ changes the relative population of metastable states separated by high barriers: probability must then be moved from one basin to another. A continuous flow can only do so by sweeping it through a region where the density is essentially zero, at a cost that grows with the barrier, and when the basins are effectively disconnected it cannot do so at all. Tellingly, the double-well benchmark used to illustrate escorted simulations in Refs.~\cite{VJ2008,LeeJarzynski2026}, Sun's model~\cite{Sun2003}, is symmetric under $q\to-q$. Its two wells always carry the same weight, and the problem never shows up.

In this Letter we lift this restriction by allowing the escort to jump. We supplement the flow with a teleportation kernel, i.e., a jump process whose rate is proportional to $\dot\lambda$; we derive the corresponding nonequilibrium identity and the condition under which flow and jumps together eliminate the lag. We then show that, whenever the barriers are not crossed during switching, every continuous escort is subject to a variance floor fixed solely by the change in basin weights, and that teleportation removes it, with a minimal number of jumps set by the same weights. An exactly solvable two-basin model and a tilted double well built on Sun's potential illustrate these results.

We denote by $\mathcal L_\lambda$ the generator of the physical dynamics (Hamiltonian, Langevin, or Monte Carlo), which preserves $p_{\rm eq}$ at fixed $\lambda$. As in Ref.~\cite{VJ2008} we add the flow $\dot\lambda\mathbf u$; in addition, we allow the system to jump. During an increment $d\lambda$, a system at $\mathbf z$ is teleported with probability $a_\lambda(\mathbf z)\,d\lambda$ to a new point $\mathbf z'$ drawn from a landing density $q_\lambda(\mathbf z'|\mathbf z)$. The phase-space density $f(\mathbf z,t)$ then evolves according to
\begin{equation}
\partial_t f=\mathcal L_\lambda f-\dot\lambda\,\nabla\!\cdot(\mathbf u f)+\dot\lambda\,\mathcal J_\lambda f ,
\label{eq:FP}
\end{equation}
\begin{equation}
(\mathcal J_\lambda f)(\mathbf z)=\int\! d\mathbf z'\,\big[k_\lambda(\mathbf z'\!\to\!\mathbf z)f(\mathbf z')-k_\lambda(\mathbf z\!\to\!\mathbf z')f(\mathbf z)\big],
\label{eq:J}
\end{equation}
\begin{equation}
k_\lambda(\mathbf z\!\to\!\mathbf z')=a_\lambda(\mathbf z)\,q_\lambda(\mathbf z'|\mathbf z),\qquad \mathcal L_\lambda e^{-\beta H_\lambda}=0 .
\label{eq:kernel}
\end{equation}
Apart from normalization, the kernel is essentially arbitrary: it may be local or global, it need not satisfy detailed balance, and it may depend on $\lambda$ in any way. Generators of this type have recently been exploited in generative modeling~\cite{GeneratorMatching2025} and in the sampling of discrete distributions~\cite{LEAPS2025}.

The derivation of an identity for $\Delta F$ follows Refs.~\cite{VJ2008,HummerSzabo2001} step by step. We define
\begin{multline}
\frac{\delta H}{\delta\lambda}\equiv\frac{\partial H}{\partial\lambda}+\mathbf u\cdot\nabla H-\beta^{-1}\nabla\!\cdot\mathbf u\\
+\beta^{-1}\Big[\!\int\! d\mathbf z'\,k_\lambda(\mathbf z'\!\to\!\mathbf z)\,e^{\beta[H(\mathbf z)-H(\mathbf z')]}-a_\lambda(\mathbf z)\Big],
\label{eq:dHdl}
\end{multline}
and verify by direct substitution that $g(\mathbf z,t)=e^{-\beta H(\mathbf z,\lambda_t)}/Z_A$ solves the sink equation $\partial_t g=\mathcal L_\lambda g-\dot\lambda\nabla\!\cdot(\mathbf u g)+\dot\lambda\,\mathcal J_\lambda g-\beta\dot\lambda\,(\delta H/\delta\lambda)\,g$. A second solution follows from the Feynman--Kac theorem, which applies to any Markov process, with or without jumps~\cite{HummerSzabo2001}. Let each trajectory of the modified dynamics, started in equilibrium at $\lambda=A$, carry the weight $e^{-\beta w_t}$, with
\begin{equation}
w_t=\int_0^t\! dt'\,\dot\lambda\,\frac{\delta H}{\delta\lambda}\big(\mathbf z_{t'},\lambda_{t'}\big).
\label{eq:wt}
\end{equation}
The weighted density $\avg{\delta(\mathbf z-\mathbf z_t)\,e^{-\beta w_t}}_{\mathbf u,k}$ then obeys the same sink equation: the first three terms move the trajectories, the last one updates their weights. Since the equation is linear and both solutions start from $p_{\rm eq}(\mathbf z,A)$ at $t=0$, they coincide,
\begin{equation}
\frac{1}{Z_A}\,e^{-\beta H(\mathbf z,\lambda_t)}=\avg{\delta(\mathbf z-\mathbf z_t)\,e^{-\beta w_t}}_{\mathbf u,k} .
\label{eq:FK}
\end{equation}
Setting $t=\tau$ and integrating over $\mathbf z$, we obtain
\begin{equation}
e^{-\beta\Delta F}=\avg{e^{-\beta W}}_{\mathbf u,k},\qquad W\equiv w_\tau .
\label{eq:identity}
\end{equation}
Compared with Ref.~\cite{VJ2008}, the new ingredient is the term in square brackets in Eq.~(\ref{eq:dHdl}), which accounts for the probability removed from $\mathbf z$ by jumps ($-a_\lambda$) and deposited there from elsewhere (the integral). Since this integral is generally not available in closed form, in practice we use an alternative work functional, obtained by comparing the probability of each trajectory with that of its time reverse, in the spirit of Crooks~\cite{Crooks1998,Crooks1999}. Introducing reverse jump rates $\tilde a_\lambda$ and landing densities $\tilde q_\lambda$, which are again arbitrary up to the usual requirement that every reverse jump can also occur in the forward direction, the work along a trajectory with jumps $\mathbf z^-\to\mathbf z^+$ is
\begin{multline}
\beta W=\beta\!\int\! d\lambda\,\Big[\partial_\lambda H+\mathbf u\cdot\nabla H-\beta^{-1}\nabla\!\cdot\mathbf u\Big]\\
-\!\int\! d\lambda\,\big[a_\lambda(\mathbf z_\lambda)-\tilde a_\lambda(\mathbf z_\lambda)\big]\\
+\sum_{\rm jumps}\Big\{\beta\big[H(\mathbf z^+)-H(\mathbf z^-)\big]\\
+\ln\frac{a(\mathbf z^-)\,q(\mathbf z^+|\mathbf z^-)}{\tilde a(\mathbf z^+)\,\tilde q(\mathbf z^-|\mathbf z^+)}\Big\},
\label{eq:Wlocal}
\end{multline}
and it satisfies the same identity, Eq.~(\ref{eq:identity}). Only the Hamiltonian and normalized landing densities are needed; no partition function enters. A derivation, which also yields an identity valid exactly at finite time step, is given in Appendix~\ref{app:local}. Without jumps, Eq.~(\ref{eq:Wlocal}) reduces to the escorted work of Ref.~\cite{VJ2008}, and for $\mathbf u=0$ to the ordinary work of Eq.~(\ref{eq:JE}); without flow, it is the continuous-time analogue of the incremental weights of sequential Monte Carlo with a backward kernel~\cite{DelMoral2006}.

As in Ref.~\cite{VJ2008}, the identity is most useful when the modified dynamics keeps the system close to equilibrium. Suppose that flow and jumps together carry $p_{\rm eq}(\cdot,\lambda)$ exactly onto $p_{\rm eq}(\cdot,\lambda+d\lambda)$, so that $f=p_{\rm eq}$ solves Eq.~(\ref{eq:FP}). Substituting this solution, we find
\begin{equation}
\partial_\lambda p_{\rm eq}+\nabla\!\cdot(\mathbf u\,p_{\rm eq})=\mathcal J_\lambda p_{\rm eq}
\;\;\Longrightarrow\;\;
\frac{\delta H}{\delta\lambda}=\frac{dF}{d\lambda},\quad W=\Delta F .
\label{eq:perfect}
\end{equation}
In other words, there is no dissipation and a single realization yields the correct free energy difference. The same holds for the trajectory form, Eq.~(\ref{eq:Wlocal}), provided the reverse jumps are those of the time-reversed process, $\tilde a_\lambda(\mathbf z')\,\tilde q_\lambda(\mathbf z|\mathbf z')=a_\lambda(\mathbf z)\,q_\lambda(\mathbf z'|\mathbf z)\,p_{\rm eq}(\mathbf z,\lambda)/p_{\rm eq}(\mathbf z',\lambda)$: with this choice the jump terms in Eq.~(\ref{eq:Wlocal}) vanish identically, $\tilde a_\lambda$ becomes the inflow term of Eq.~(\ref{eq:dHdl}), and the two work functionals coincide trajectory by trajectory. Equation~(\ref{eq:perfect}) is the analogue, with jumps, of Eq.~(15) of Ref.~\cite{VJ2008}. Its right-hand side acts as a nonlocal source term: it removes probability at one point and creates it at another, which no continuity equation can do.

Why should jumps be needed at all? Consider a system whose configuration space contains metastable basins $\Omega_c$, separated by barriers that are not crossed on the time scale $\tau$ of the simulation, and let
\begin{equation}
\pi_c(\lambda)=\int_{\Omega_c}\! d\mathbf z\,p_{\rm eq}(\mathbf z,\lambda)
\label{eq:weights}
\end{equation}
be the equilibrium weight of basin $c$; we write $\pi^A$ and $\pi^B$ for the weights at the two end points. A continuous flow field generates a continuous, invertible map of phase space. It could move probability between basins only by dragging it through the barrier region, where the equilibrium density is vanishingly small; excluding this, every escorted trajectory ends in the basin where it started, and the fraction of trajectories in basin $c$ stays equal to $\pi_c(A)$ throughout the process. If the weights change with $\lambda$, the lag cannot be removed, whatever $\mathbf u$ we choose.

This intuition can be made quantitative. Let $\hat\pi_c$ be the fraction of trajectories that end in basin $c$, for any dynamics, with or without flow and jumps. Integrating Eq.~(\ref{eq:FK}) at $t=\tau$ over $\Omega_c$ gives $\hat\pi_c\,\avg{e^{-\beta W}\,|\,\mathbf z_\tau\in\Omega_c}=e^{-\beta\Delta F}\,\pi_c(B)$: the trajectories that end in basin $c$ must carry, on average, a weight proportional to $\pi_c(B)/\hat\pi_c$. (The same relation holds for the trajectory form of the work, Eq.~(\ref{eq:Wlocal}), since the reverse process starts in equilibrium at $\lambda=B$; see Appendix~\ref{app:floors}.) Applying Jensen's inequality within each basin, we find
\begin{equation}
\frac{{\rm Var}\big[e^{-\beta W}\big]}{\avg{e^{-\beta W}}^2}\;\ge\;\chi^2\big(\pi^B\big\|\hat\pi\big)=\sum_c\frac{\pi_c(B)^2}{\hat\pi_c}-1 .
\label{eq:chi2general}
\end{equation}
This is the coarse-grained version of a familiar fact about importance sampling: the relative variance of the weights cannot be smaller than the $\chi^2$ divergence between the target and the distribution actually sampled at the final time~\cite{Doucet2022}. For a continuous escort without barrier crossings $\hat\pi=\pi^A$, and Eq.~(\ref{eq:chi2general}), together with the analogous bound on the mean work obtained by applying Jensen's inequality to $W$ itself, yields
\begin{equation}
\begin{gathered}
\frac{{\rm Var}\big[e^{-\beta W}\big]}{\avg{e^{-\beta W}}^2}\ge\chi^2\big(\pi^B\big\|\pi^A\big),\\
\beta\avg{W}-\beta\Delta F\ge D_{\rm KL}\big(\pi^A\big\|\pi^B\big).
\end{gathered}
\label{eq:floor}
\end{equation}
These floors, derived in Appendix~\ref{app:floors}, depend only on how the basin weights change, not on $\mathbf u$: no amount of ingenuity in the design of the flow can beat them. They are also easy to evaluate. When a basin that is initially almost empty must end up holding a finite fraction of the probability, $\pi_c(B)^2/\pi_c(A)$ is huge, and so is the number of trajectories needed.

Jumps lift the constraint, because a single jump can carry a trajectory from one basin to another. Since each such jump transfers the probability of one trajectory, the mean number $\avg{n}$ of inter-basin jumps per trajectory obeys
\begin{gather}
\avg{n}\;\ge\;\tfrac12\sum_c\big|\hat\pi_c-\pi_c(A)\big|,
\label{eq:activity}\\
{\rm Var}\big[e^{-\beta W}\big]=0\;\Rightarrow\;\avg{n}\ge d_{\rm TV}\big(\pi^A,\pi^B\big),
\label{eq:threshold}
\end{gather}
where the second line follows from Eq.~(\ref{eq:chi2general}), which requires $\hat\pi=\pi^B$ for zero variance, and $d_{\rm TV}$ is the total variation distance. A zero-variance estimator thus needs a minimal teleportation activity, fixed once again by the basin weights alone, in the spirit of thermodynamic speed limits involving dynamical activity~\cite{ShiraishiFunoSaito2018}.

These bounds are saturated by a model simple enough to be solved by hand. Take two identical, disconnected basins, $L$ and $R$, and switch off an energy offset on the right one as $\lambda$ goes from $A=0$ to $B=1$:
\begin{equation}
\begin{gathered}
H_\lambda(\mathbf z)=H_0(\mathbf z)+(1-\lambda)\,\Delta\;\mathbf 1_{R}(\mathbf z),\\
\pi_R(\lambda)=\frac{1}{1+e^{\beta(1-\lambda)\Delta}} .
\end{gathered}
\label{eq:toy}
\end{equation}
The right basin starts with weight $\pi_R(0)\simeq e^{-\beta\Delta}$ and ends with $\pi_R(1)=1/2$, while the shape of each basin never changes. The best escort is therefore $\mathbf u=0$, and the work is $W=0$ for trajectories in $L$ and $W=-\Delta$ for those in $R$. The resulting variance saturates the floor~(\ref{eq:floor}),
\begin{equation}
\frac{{\rm Var}\big[e^{-\beta W}\big]}{\avg{e^{-\beta W}}^2}\bigg|_{\rm escort}=\sinh^2\!\frac{\beta\Delta}{2},
\label{eq:toyescort}
\end{equation}
which grows as $e^{\beta\Delta}/4$: the estimate rests entirely on the rare trajectories that happen to start on the right. Now add teleports from $L$ to $R$, landing on the equilibrium distribution of the right basin,
\begin{equation}
\begin{gathered}
a_\lambda(\mathbf z)=\beta\Delta\,\pi_R(\lambda)\,\mathbf 1_L(\mathbf z),\\
q_\lambda(\mathbf z'|\mathbf z)=\frac{p_{\rm eq}(\mathbf z',\lambda)\,\mathbf 1_R(\mathbf z')}{\pi_R(\lambda)} .
\end{gathered}
\label{eq:toykernel}
\end{equation}
The rate is chosen so that the flux of jumps, $a_\lambda\pi_L$, equals $d\pi_R/d\lambda=\beta\Delta\,\pi_L\pi_R$. One checks directly that Eq.~(\ref{eq:perfect}) is satisfied, and hence
\begin{gather}
W=\Delta F\quad\text{for every trajectory},
\label{eq:toyresult}\\
\avg{n}=\pi_R(1)-\pi_R(0)=d_{\rm TV}\big(\pi^A,\pi^B\big).
\label{eq:toyn}
\end{gather}
The teleport removes the floor completely, and it does so with the smallest number of jumps allowed by Eq.~(\ref{eq:threshold}). The result is also robust. Rescaling the rate to $c\,a_\lambda$, so that the kernel is no longer perfect, we computed the variance exactly with the work functional of Eq.~(\ref{eq:wt})~\cite{SM}: for $\beta\Delta=10$, where the escort gives $5.5\times10^3$, it is $2.1$ for $c=1/2$ and $0.55$ for $c=2$. Teleports that are wrong by a factor of two still reduce the variance by more than three orders of magnitude.

\begin{figure}[t]
\includegraphics[width=\columnwidth]{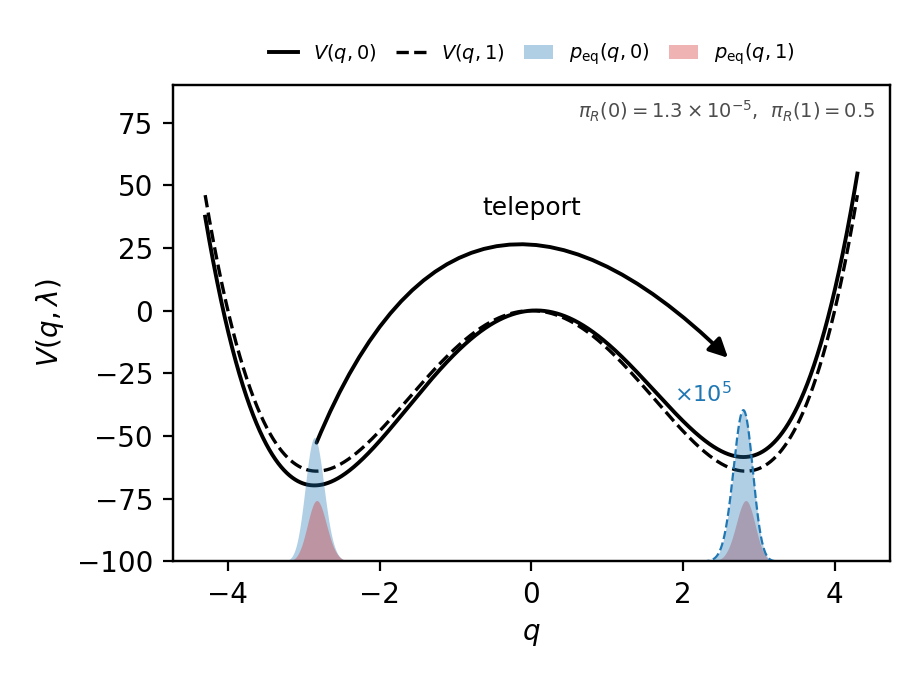}
\caption{Tilted double well, Eq.~(\ref{eq:sun}), with $h=2$, at $\lambda=0$ (solid) and $\lambda=1$ (dashed), with the corresponding equilibrium distributions (shaded). The weight of the right well grows from $1.3\times10^{-5}$ to $1/2$; its distribution at $\lambda=0$ is shown magnified $10^5$ times. The tilt also shifts the minima slightly, which the escort field compensates. The arrow indicates teleportation.}
\label{fig:landscape}
\end{figure}

We now turn to a continuous model in the spirit of Ref.~\cite{VJ2008}. We freeze Sun's double well at $\lambda=0$ and add a linear tilt that is switched off,
\begin{equation}
V(q,\lambda)=q^4-16q^2+(1-\lambda)\,h\,q,\qquad \lambda:0\to1 ,
\label{eq:sun}
\end{equation}
with $\beta=1$. The barrier between the wells stays above $55\,k_BT$ throughout, so the wells remain dynamically disconnected. The tilt moves the minima $q_\pm(\lambda)$ only slightly, but it changes the weights of the wells by orders of magnitude: for $h=2$ the right well holds a fraction $\pi_R(0)=1.3\times10^{-5}$ of the probability at $\lambda=0$ and half of it at $\lambda=1$ (Fig.~\ref{fig:landscape}). Here $\Delta F=4.978$, and Eq.~(\ref{eq:floor}) sets a floor $\chi^2=2.0\times10^4$ for any escort. Unlike Sun's original model, in which the wells merge at $\lambda=1$ and a smooth flow can in principle do the job, here a flow could transfer probability only by dragging it across a barrier of more than $55\,k_BT$.

We simulate overdamped Langevin dynamics with Metropolis-adjusted steps, so that the identity holds exactly at finite time step (Appendix~\ref{app:local}), and vary $\lambda$ linearly in a time $\tau$. For the escort we use, in the spirit of Eq.~(19) of Ref.~\cite{VJ2008}, a field that moves each well along with its minimum and interpolates smoothly across the barrier over a width $w=1/2$ (dots denote derivatives with respect to $\lambda$),
\begin{equation}
\begin{gathered}
u(q,\lambda)=\dot q_-+\big(\dot q_+-\dot q_-\big)\frac{1+\tanh(q/w)}{2},\\
\dot q_\pm\equiv\frac{dq_\pm}{d\lambda}=\frac{h}{V''(q_\pm)} .
\end{gathered}
\label{eq:sunescort}
\end{equation}
For the teleport we deliberately use crude ingredients that require no knowledge of $\Delta F$. Forward jumps go from the left well to the right one and reverse jumps the other way, with rates set by harmonic estimates of the basin weights, and landing densities given by the harmonic approximation of each well,
\begin{gather}
a_\lambda=\frac{\dot\pi^{\rm h}_R}{\pi^{\rm h}_L}\,\mathbf 1_{q<0},\qquad
\tilde a_\lambda=\frac{\dot\pi^{\rm h}_R}{\pi^{\rm h}_R}\,\mathbf 1_{q>0},
\label{eq:sunrates}\\
q_\lambda=\mathcal N\!\Big(q_+,\frac{1}{\beta V''(q_+)}\Big),\qquad
\tilde q_\lambda=\mathcal N\!\Big(q_-,\frac{1}{\beta V''(q_-)}\Big),
\label{eq:sunkernel}
\end{gather}
\begin{equation}
\pi^{\rm h}_R=\frac{Z^{\rm h}_+}{Z^{\rm h}_-+Z^{\rm h}_+},\qquad
Z^{\rm h}_\pm=e^{-\beta V(q_\pm,\lambda)}\sqrt{\frac{2\pi}{\beta V''(q_\pm)}} .
\label{eq:harmonic}
\end{equation}

\begin{figure}[t]
\includegraphics[width=\columnwidth]{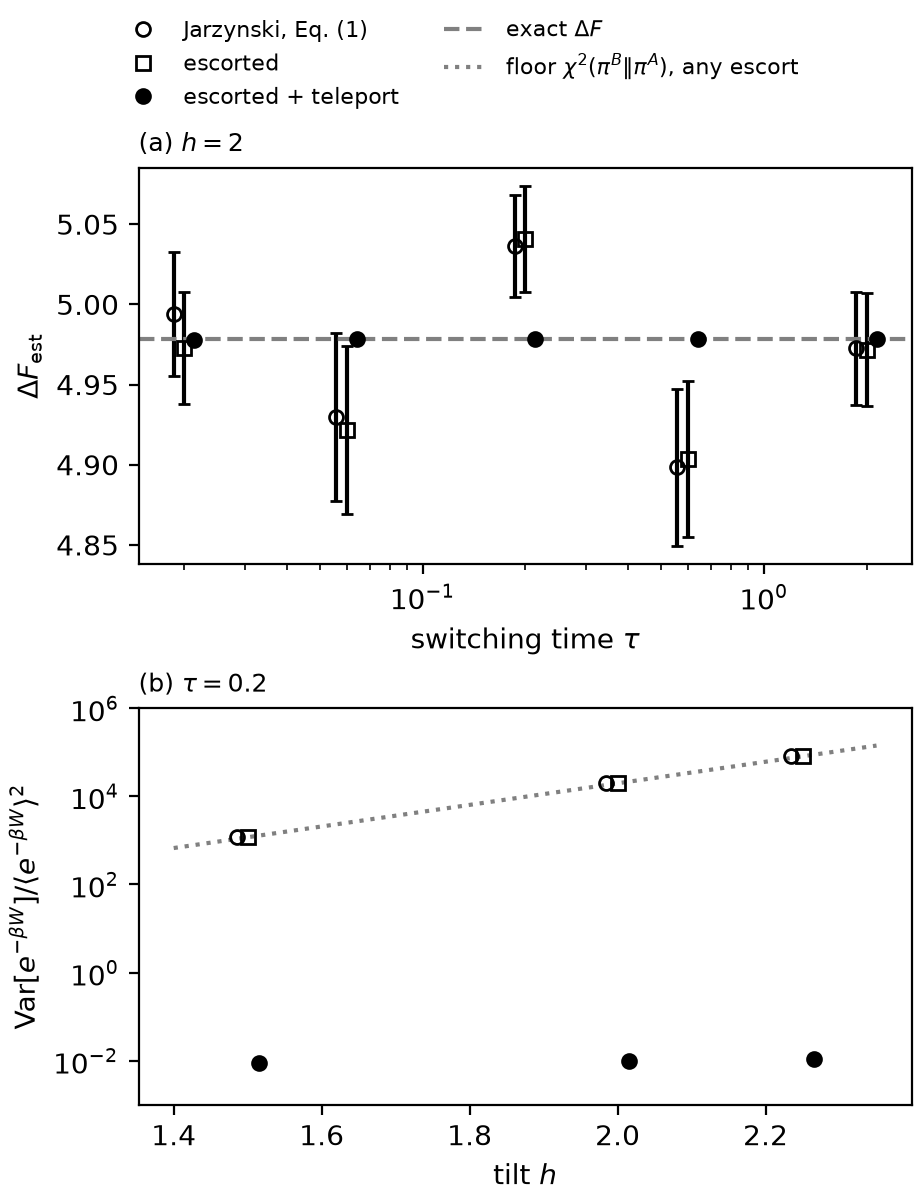}
\caption{(a) Free energy estimates versus switching time $\tau$ for the model of Fig.~\ref{fig:landscape} ($h=2$, $N=10^7$ trajectories per point): Eq.~(\ref{eq:JE}) (open circles), escorted simulations (open squares), and escorted simulations with teleportation (filled circles); dashed line, exact $\Delta F$. Error bars are standard errors over 20 batches; for teleported simulations they are smaller than the symbols. (b) Measured relative variance of $e^{-\beta W}$ versus tilt $h$ at $\tau=0.2$, with $N=10^6$, $10^7$, and $3\times10^7$ trajectories for $h=1.5$, $2$, and $2.25$, so that about $10^2$ trajectories start in the right well; dotted line, the floor of Eq.~(\ref{eq:floor}), valid for any escort.}
\label{fig:estimates}
\end{figure}

Figure~\ref{fig:estimates}(a) shows the estimates for $h=2$, with $N=10^7$ trajectories per point for all three methods. About a hundred of them, $N\pi_R(0)\simeq130$, start in the right well, so the rare events that matter are sampled and all estimates are correct on average. The difference lies in the fluctuations. Without teleports the error bars are about $0.04$ and do not shrink with $\tau$: the measured relative variance of $e^{-\beta W}$ is $2\times10^4$ at every switching time, with and without the escort, right on the floor of Eq.~(\ref{eq:floor}). Slower switching does not help, because the barrier is never crossed, and no escort can help either. With teleports the variance drops to between $0.3$, at the shortest $\tau$, and $8\times10^{-3}$, and the error bars shrink by more than two orders of magnitude; at the longer switching times, an escorted simulation would need more than $10^6$ times as many trajectories to reach the same precision. Figure~\ref{fig:estimates}(b) shows that none of this depends on the chosen tilt: as $h$ grows, the floor rises exponentially and the variance of the plain and escorted estimates follows it, while with teleports it stays around $10^{-2}$.

When the small well is so rare that no trajectory starts there, the failure becomes systematic. For $h=3$, $\pi_R(0)=4.5\times10^{-8}$, and with $N=10^6$ trajectories the plain and escorted estimates return, at every $\tau$,
\begin{equation}
\Delta F_{\rm est}\;\longrightarrow\;\Delta F-\beta^{-1}\ln\pi_L(1)=\Delta F+\beta^{-1}\ln 2 ,
\label{eq:ln2}
\end{equation}
with small but misleading error bars, as follows from the basin relation that leads to Eq.~(\ref{eq:chi2general}) with no trajectory in the right well; teleported simulations give the exact value~\cite{SM}. The intermediate regime of Fig.~\ref{fig:estimates} shows that this is not a sampling artifact: whenever the rare trajectories are present, the variance of an escorted estimate sits exactly where Eq.~(\ref{eq:floor}) puts it.

As in Ref.~\cite{VJ2008}, the key to success is a modified dynamics that mimics the effect of $\lambda$ on $p_{\rm eq}$: escorting when $\lambda$ deforms the equilibrium distribution, teleportation when $\lambda$ redistributes probability among states that the dynamics cannot connect. The price is that landing densities must be normalized and should resemble the basins they target, much like a proposal in importance sampling. Crude harmonic approximations were enough here; in molecular systems they could come from short equilibrium runs in each basin or from learned generative models~\cite{Wirnsberger2020,GeneratorMatching2025}. A discontinuous invertible map, admissible in discrete-time escorts~\cite{Jarzynski2002,VJ2011}, can also move probability between basins, but it requires knowing exactly which part of one basin should be mapped onto another.

Jump moves are familiar in sampling, from annealed importance sampling and sequential Monte Carlo~\cite{Neal2001,DelMoral2006} to recent schemes that combine transport with jumps or nonlocal proposals~\cite{LEAPS2025,FKC2025,Schonle2025,Schonle2025b,LeeSantanaGijzen2025}, and the imbalance of mass between modes is known to control the complexity of annealing~\cite{Guo2026,HanIyerSlepcev2026,Agarwal2026}. Here we have brought this mechanism into escorted free energy simulations, and identified an escort-independent variance floor together with the minimal teleportation activity needed to remove it. Basin weights could also be estimated separately by reweighting~\cite{Monmarche2026}; teleportation folds this step into a single nonequilibrium identity, while the escort handles the deformation within each basin.

Teleportation should be most useful against hidden barriers, for instance when the populations of side-chain rotamers or ligand binding poses differ between the end states of an alchemical transformation, and in discrete systems, where jumps are the only possible escort. It could also help in training generative AI models defined by an energy function: there, each training step requires samples of a model Boltzmann distribution that keeps changing, and Jarzynski weights are used to correct walkers that cannot follow the shifting weights of its modes~\cite{Carbone2023,Carbone2025}; teleports would let the walkers follow them. As for the original escort~\cite{VJ2008}, the approach extends directly to Crooks's fluctuation theorem~\cite{Crooks1999} and, through Eq.~(\ref{eq:FK}), to the Hummer--Szabo identity for potentials of mean force~\cite{HummerSzabo2001}.

\textit{Data availability}---The code and data that support the findings of this article are openly available in the Supplemental Material.\\
\textit{IA Statement}---
The author used an AI assistant (Claude, Anthropic) to help polish the text, write and test the simulation code. All content was reviewed and verified by the author, who takes full responsibility for it.

\bibliography{refs}

\section*{End Matter}
\appendix

\section{Trajectory form of the work}
\label{app:local}
We derive Eq.~(\ref{eq:Wlocal}) as the continuum limit of a discrete-time scheme for which the identity holds exactly at any step size; the same scheme is used in the simulations~\cite{SM}. Divide the protocol into $K$ steps $\lambda_0=A<\dots<\lambda_K=B$ and write $\gamma_k=e^{-\beta H(\cdot,\lambda_k)}$, $Z_k=\int\gamma_k$. Starting from $\mathbf z_0\sim\gamma_0/Z_0$, step $k\to k+1$ consists of (i) an escort map $\mathbf y_k=M_k(\mathbf z_k)$, a bijection with Jacobian $J_k$, together with $\lambda_k\to\lambda_{k+1}$; (ii) a teleport move $\mathbf x_k\sim\mathcal K_{k+1}(\mathbf y_k\to\cdot)$; (iii) a move $\mathbf z_{k+1}\sim\mathcal P_{k+1}(\mathbf x_k\to\cdot)$ of the physical dynamics, which leaves $\gamma_{k+1}$ invariant. With an arbitrary normalized backward kernel $\mathcal B_{k+1}$, such that $\mathcal K_{k+1}(\mathbf y\to\mathbf x)>0$ whenever $\gamma_{k+1}(\mathbf y)\mathcal B_{k+1}(\mathbf x\to\mathbf y)>0$, we define
\begin{equation}
e^{-\beta W}=\prod_{k=0}^{K-1}\frac{\gamma_{k+1}(\mathbf y_k)\,J_k(\mathbf z_k)}{\gamma_k(\mathbf z_k)}\,
\frac{\gamma_{k+1}(\mathbf x_k)\,\mathcal B_{k+1}(\mathbf x_k\to\mathbf y_k)}{\gamma_{k+1}(\mathbf y_k)\,\mathcal K_{k+1}(\mathbf y_k\to\mathbf x_k)} .
\label{eq:factors}
\end{equation}
Let $w_k$ be the product of the first $k$ factors. By induction, $\mathbb E[w_k f(\mathbf z_k)]=Z_0^{-1}\int\gamma_k f$ for any test function $f$: the first factor turns $\gamma_k$ into $\gamma_{k+1}$ through the change of variables $\mathbf y=M_k(\mathbf z)$; the second leaves $\gamma_{k+1}$ unchanged because $\int d\mathbf y\,\mathcal B_{k+1}(\mathbf x\to\mathbf y)=1$; and move (iii) preserves $\gamma_{k+1}$ by assumption. At $k=K$,
\begin{equation}
\begin{gathered}
\avg{e^{-\beta W}}=e^{-\beta\Delta F},\\
\avg{\delta(\mathbf z-\mathbf z_K)\,e^{-\beta W}}=e^{-\beta\Delta F}p_{\rm eq}(\mathbf z,B),
\end{gathered}
\label{eq:exact}
\end{equation}
exactly, for any $K$; only invariance under $\mathcal P_{k+1}$ is needed, not detailed balance. For $\delta\lambda=\lambda_{k+1}-\lambda_k\to0$, take $M_k(\mathbf z)=\mathbf z+\mathbf u\,\delta\lambda$, so that $\ln J_k=\nabla\!\cdot\mathbf u\,\delta\lambda$, and
\begin{equation}
\begin{gathered}
\mathcal K(\mathbf y\to\mathbf x)=(1-a\,\delta\lambda)\,\delta(\mathbf x-\mathbf y)+a(\mathbf y)\,\delta\lambda\,q(\mathbf x|\mathbf y),\\
\mathcal B(\mathbf x\to\mathbf y)=(1-\tilde a\,\delta\lambda)\,\delta(\mathbf y-\mathbf x)+\tilde a(\mathbf x)\,\delta\lambda\,\tilde q(\mathbf y|\mathbf x).
\end{gathered}
\label{eq:KB}
\end{equation}
The first factor in Eq.~(\ref{eq:factors}) then contributes $\beta[\partial_\lambda H+\mathbf u\cdot\nabla H-\beta^{-1}\nabla\!\cdot\mathbf u]\,\delta\lambda$ to $\beta W$. If no jump occurs, only the singular parts of Eq.~(\ref{eq:KB}) enter, and the second factor contributes $-(a-\tilde a)\,\delta\lambda$; if a jump $\mathbf y\to\mathbf x$ occurs, only the regular parts enter, and it contributes $\beta[H(\mathbf x)-H(\mathbf y)]+\ln[a\,q(\mathbf x|\mathbf y)/\tilde a\,\tilde q(\mathbf y|\mathbf x)]$. Summing over steps gives Eq.~(\ref{eq:Wlocal}).

If the backward jumps are the time reversal of the forward ones, $\tilde a(\mathbf x)\tilde q(\mathbf y|\mathbf x)=a(\mathbf y)q(\mathbf x|\mathbf y)\,p_{\rm eq}(\mathbf y)/p_{\rm eq}(\mathbf x)$, every jump term vanishes, and integrating over $\mathbf y$ shows that $\tilde a(\mathbf x)$ equals the inflow term of Eq.~(\ref{eq:dHdl}). Equation~(\ref{eq:Wlocal}) then reduces to Eq.~(\ref{eq:wt}) trajectory by trajectory.

\section{Variance and dissipation floors}
\label{app:floors}
Let $\hat\pi_c$ be the probability that a trajectory ends in basin $\Omega_c$ and $\avg{\,\cdot\,|\,c}$ the corresponding conditional average. Integrating the second relation in Eq.~(\ref{eq:exact}), or Eq.~(\ref{eq:FK}) at $t=\tau$, over $\Omega_c$ gives
\begin{equation}
\hat\pi_c\,\avg{e^{-\beta W}\,|\,c}=e^{-\beta\Delta F}\,\pi_c(B),
\label{eq:basin}
\end{equation}
for any dynamics and either form of the work; in particular $\hat\pi_c>0$ whenever $\pi_c(B)>0$. Decomposing the second moment over final basins and using $\avg{X^2|c}\ge\avg{X|c}^2$,
\begin{equation}
\avg{e^{-2\beta W}}\ge\sum_c\hat\pi_c\,\avg{e^{-\beta W}|c}^2=e^{-2\beta\Delta F}\sum_c\frac{\pi_c(B)^2}{\hat\pi_c},
\end{equation}
which, divided by $\avg{e^{-\beta W}}^2=e^{-2\beta\Delta F}$, is Eq.~(\ref{eq:chi2general}). Equality holds if and only if $e^{-\beta W}$ is constant within each final basin. Similarly, Jensen's inequality $\avg{e^{-\beta W}|c}\ge e^{-\beta\avg{W|c}}$ and Eq.~(\ref{eq:basin}) give $\avg{W|c}\ge\Delta F-\beta^{-1}\ln[\pi_c(B)/\hat\pi_c]$, hence
\begin{equation}
\beta\avg{W}-\beta\Delta F\ge\sum_c\hat\pi_c\ln\frac{\hat\pi_c}{\pi_c(B)}=D_{\rm KL}\big(\hat\pi\big\|\pi^B\big).
\end{equation}
A continuous escort generates a homeomorphism of phase space at each step; if neither the flow nor the physical dynamics carries trajectories across the barriers, every trajectory ends in its initial basin, $\hat\pi=\pi^A$, and Eq.~(\ref{eq:floor}) follows. Without such crossings, basin occupations change only through inter-basin jumps. With $n_c^{\rm in}$ and $n_c^{\rm out}$ the numbers of jumps into and out of basin $c$ and $n=\frac12\sum_c(n_c^{\rm in}+n_c^{\rm out})$, we have $\hat\pi_c-\pi_c(A)=\avg{n_c^{\rm in}}-\avg{n_c^{\rm out}}$, so that $\frac12\sum_c|\hat\pi_c-\pi_c(A)|\le\avg{n}$, which is Eq.~(\ref{eq:activity}). Zero variance requires $\hat\pi=\pi^B$, which gives Eq.~(\ref{eq:threshold}).

\end{document}